\documentclass[%
 reprint,
showpacs,
 amsmath,amssymb,
 prl
]{revtex4-1}

\usepackage{graphicx}
\usepackage{dcolumn}
\usepackage{bm}
\usepackage{threeparttable}
\newcommand\aap{Astronomy and Astrophysics}
\newcommand\apjl{ApJL}
\newcommand\mnras{MNRAS}

\begin{document}

\preprint{APS/123-QED}

\title{High-energy neutrino flare from cloud-jet interaction in the blazar PKS 0502+049}

\author{Hao-Ning He}
\affiliation{Key Laboratory of Dark Matter and Space Astronomy, Purple Mountain Observatory, Chinese Academy of Sciences, Nanjing 210034, China}
\affiliation{Astrophysical Big Bang Laboratory, RIKEN, Wako, Saitama, Japan}

\author{Yoshiyuki Inoue}
\author{Susumu Inoue}
\affiliation{Interdisciplinary Theoretical \& Mathematical Science Program (iTHEMS),\\
RIKEN, 2-1 Wako, Saitama 351-0198, Japan}

\author{Yun-Feng Liang}
\affiliation{Key Laboratory of Dark Matter and Space Astronomy, Purple Mountain Observatory, Chinese Academy of Sciences, Nanjing 210034, China}

\date{\today}

\begin{abstract}
Following the detection of a $\sim$300 TeV neutrino potentially associated with the flaring blazar TXS~0506+056, an excess of neutrinos around its position in 2014-2015 was revealed by IceCube. However, its contemporaneous quiescence in $\gamma$-rays is challenging to interpret consistently. Meanwhile, the blazar PKS~0502+049, positioned within the neutrino localization uncertainties, was seen to be flaring in $\gamma$-rays. We show that dense, line-emitting gas clouds that interact with its jet and induce cosmic ray acceleration and hadronuclear interaction can plausibly explain the 2014-2015 neutrino flare. 
\end{abstract}
\maketitle

{\it Introduction}
The IceCube Observatory detected a $\sim300$~TeV neutrino on September 22, 2017 that was spatially and temporally coincident with a $\gamma$-ray flare from the blazar TXS 0506+056, with a significance of $\sim 3\sigma$ \citep{IceCube2018MultiMessenger}.
The potential association between the activity of TXS 0506+056 and the event makes the object a promising candidate source of high-energy neutrinos.
Blazars are active galactic nuclei powered by supermassive black holes (SMBHs) with relativistic jets pointing nearly toward the observer.
The neutrino can be generated through acceleration of protons to sufficiently high energies in the jet and its subsequent photopion interaction with ambient radiation fields, either internal or external to the jet \citep{Murase2018blazarneutrino,Righi2018blazarneutrino,Gao2018blazarneutrino,Ansoldi2018blazarneutrino}.
An alternative possibility is hadronuclear interactions of accelerated protons with dense gas clouds in the vicinity of the jet \citep{Liu2018blazarneutrino}, although there is no observational evidence in TXS~0506+056 of broad line emission that point to the presence of such clouds \citep{Paiano2018}.

Furthermore, the IceCube Collaboration investigated 9.5 years of historical data around the position of the blazar.
They also found an excess of events with a significance of $3.5\sigma$, consisting of $13\pm5$ neutrinos with energy above $30$~TeV during a 158-day box-shaped time window from MJD 56937.81 to MJD 57096.21 (September 2014 to March 2015), which is above the expected atmospheric background \citep{IceCube2018neutrinoflare}.

During this time window, the blazar TXS 0506+056 was relatively inactive in $\gamma$-rays \citep{Liang2018,Padovani2018}.
The quiescent $\gamma$-ray activity is difficult to interpret consistently with the high-energy neutrino flare in 2014-2015, using the same models that were discussed for IceCube-170922A associated with the $\gamma$-ray flaring state. 
 
On the other hand, separated in the sky by $1.2^\circ$ from TXS 0506+056 is the blazar PKS 0502+049, at right ascension (RA) $76.35^\circ$, declination (Dec) $+5.00^\circ$ (J2000 equinox) \citep{Gaia2018}, and redshift $z=0.954$ \citep{Drinkwater1997}.
The latter was observed to be active in GeV $\gamma$-rays around the 158-day box-shaped time window \citep{Liang2018,Padovani2018}. The spatial coincidence between the 2014-2015 neutrino flare and PKS~0502+049 cannot be completely excluded, since the median angular resolution of the IceCube Observatory for $\sim$30 TeV neutrinos is about $0.5^\circ$ \citep{IceCube2018neutrinoflare}.

The {\it Fermi}-LAT light curve of PKS 0502+049 reveals two bright flares from MJD 56850 to MJD 57150 (i.e., from 2014-2015), overlapping with the 158-day time window, with flux about one order of magnitude higher than TXS 0506+056 \citep{Liang2018}.
Its photon flux in the energy range 0.1-500 GeV during this period is $\sim 5.0\times 10^{-7}{\rm ~ph\,cm^{-2} s^{-1}}$, much larger than that averaged over its quiescent state from MJD 56000 to MJD 56700 with $8.3\times 10^{-8}{\rm ~ph\, cm^{-2} s^{-1}}$. 
If one assumes that the $\gamma$-rays in the quiescent state has a predominantly leptonic origin (i.e. are emitted by accelerated electrons) that has not significantly varied over time, the $\gamma$-rays in the active state can be considered to be dominated by hadronic processes.
In hadronuclear interactions, the ratio between the observable flux of $\gamma$-rays to muon neutrinos is $N_{\gamma}:N_{\nu_\mu}\sim 2:1$ \citep{Kelner2006pp}, considering the equipartition among the three neutrino flavors after their oscillations during propagation.
Assuming a power-law spectrum with index -2 for the accelerated protons, the muon neutrino flux at $100~{\rm TeV}$ is estimated to be $E_{\nu_\mu}^2{dN_{\nu_\mu}}/{dE_{\nu_\mu}}\sim2.5\times10^{-8}{\rm ~GeV ~cm^{-2} ~s^{-1}}$, by extrapolating the observed $\gamma$-ray flux at 0.3~GeV.
Approximating the effective area of IceCube as $A_{\rm eff} \sim 100~{\rm m^2}$ for $\sim$100 TeV neutrinos, the number of muon neutrino events detectable by IceCube during the active phase of PKS~0502+049 is estimated to be $\sim$6, comparable to the observed number of excess events $13\pm5$.
Therefore, it is plausible that the active blazar PKS~0502+049 contributed to the neutrino flare.

Moreover, the blazar PKS~0502+049 is classified as a flat spectrum radio quasar (FSRQ) that exhibits broad emission lines, a signature of the presence of broad line region (BLR) clouds within parsecs from the SMBH.
Therefore, hadronuclear interactions between protons accelerated in the jet and dense cloud gas in PKS~0502+049 can naturally occur once the clouds enter the jet, and then produce $\gamma$-ray flares simultaneously with neutrino flares
\citep{Dar1997,Araudo2010,Barkov2012JetCloudinteraction,Ramon2012}. 

In this work, we consider hadronuclear processes induced by the interaction between the jet flow and clouds in PKS~0502+049, and calculate the expected muon neutrino detection rates by IceCube, assuming the observed $\gamma$-ray flux during its flare is due to pion decay.

{\it Cloud-Jet Interaction Model:}
The velocity dispersion of clouds in PKS~0502+049 is measured to be $v_{\rm c}=4184{\rm km/s}$ \citep{Oshlack2002BHmass}.
The moving clouds can penetrate into the jet, give rise to shocks in the jet flow behind the cloud that accelerates protons, which can then interact with the matter in the dense cloud and produce flares in both high-energy neutrinos and $\gamma$-rays \citep{Dar1997,Araudo2010,Barkov2012JetCloudinteraction,Ramon2012}. 
The crossing timescale of the cloud over the jet is estimated as
\begin{equation}
t_{\rm cr}=2 R_{\rm jet}/v_{\rm c}=166{\rm ~day~}R_{\rm jet,15.5}, 
\end{equation}
where $R_{\rm jet}=\theta z_{\rm jet}$ is the jet radius at the cloud-jet interaction site, with $\theta$ is the jet semi-opening angle and $z_{\rm jet}$ is the distance of the cloud-jet interaction site to the central SMBH. 
The whole duration of the flaring activity is about $t_{\rm act}^{\rm obs}\simeq(1+z)t_{\rm cr}$.
Given the observed duration of activity of $t_{\rm act}^{\rm obs}\sim 300$ days, the jet radius at the cloud-jet interaction site is about $R_{\rm jet}\sim 10^{15.5}~{\rm cm}$. 

Based on broadband spectral fitting of blazars, the magnetic field strength in the jet $B$ is typically estimated to be of order 0.1--1~G \citep{Inoue1996,Ghisellini2010,Inoue2016}.
Assuming a similar value for the shocked jet region, the acceleration timescale of protons can be expressed as
\begin{equation}
t_{\rm acc}\approx\frac{\epsilon E_{\rm p}}{qBc}\simeq 1\times 10^2 {\rm ~s~}\epsilon E_{\rm p,15}B_{0}^{-1},
\end{equation}
where $\epsilon$ is the acceleration efficiency, $q$ is the particle charge, and $E_{\rm p}$ is the proton energy.
For simplicity, here we set $\epsilon=1$.
Comparing the acceleration timescale with the dynamical timescale $t_{\rm dyn}=R_{\rm c}/c=3\times10^3{~\rm s~}R_{\rm c,14}$ with $R_{\rm c}$ as the size of the cloud, one can estimate the maximum energy of the protons as
\begin{equation}
E_{\rm p,max}\approx qBR_c/\epsilon\simeq3\times 10^{16}{~\rm eV~}R_{\rm c,14}B_{0}.
\label{E_pmax}
\end{equation}

Taking the average hydrogen column density of the cloud as $N_{\rm H}=10^{23}~{\rm cm^{-2}}$ \citep{Stern2014nH}, the hydrogen volume density is approximately $n_{\rm H}=10^{\rm 9}{~\rm cm^{-3}~}N_{\rm H,23}R_{\rm c,14}^{-1}$.
The characteristic cooling timescale of protons in $pp$ collisions is 
\begin{equation}
t_{\rm pp}\approx \frac{1}{\kappa_{\rm pp}\sigma_{\rm pp}n_{\rm H}c}=2\times 10^{6}{\rm ~s~} N_{\rm H,23}^{-1}R_{\rm c,14},
\end{equation}
where the $pp$ cross section and inelasticity are $\sigma_{\rm pp}=4\times10^{-26}{\rm ~cm}$ and $\kappa_{\rm pp}=0.45$ \citep{Gaisser1990}, respectively.
Since the magnetic field in the cloud is $\sim 10~{\rm G}$ \citep{Silantev2013AGNmagneticfield}, accelerated protons impinging into the cloud can be trapped within during the crossing time of the cloud \citep{Barkov2010}. 
The efficiency of $pp$ collisions can be calculated by comparing the cooling timescale and the crossing timescale, i.e., 
\begin{equation}
f_{\rm pp}={\rm min}(1,t_{\rm cr}/t_{\rm pp})\sim 1,
\end{equation}
since $t_{\rm cr}\gg t_{\rm pp}$.

The fraction of the kinetic energy of the jet channeled into protons, and the fraction of the accelerated protons that penetrate into the cloud are described as $f_{\rm dis}$ and $f_{\rm c}$, respectively.
Here we assume that most of the accelerated protons can reach the cloud, i.e., $f_{\rm c}\sim 1$，which might be achieved by a cluster of clouds with size of $R_{\rm c}\ll R_{\rm jet}$.
In hadronuclear interactions, charged and neutral pions are generated with ratio $\pi^\pm:\pi^0\approx2:1$. $\gamma$-rays are produced via the decay $\pi^0\rightarrow 2\gamma$, and the resulting $\gamma$-ray energy flux is
 \begin{equation}\label{F_gamma}
 \begin{split}
E_{\gamma}^2\frac{dN_{\gamma}}{dE_{\gamma}}&\approx\frac{1}{3}f_{\rm pp}f_{\rm dis}f_{\rm c}P_{\rm jet}/(4\pi D_{L}^2f_{\rm d})\\
& \simeq 2.5\times10^{-8}{\rm ~GeV ~cm^{-2} ~s^{-1}}~f_{\rm dis}P_{\rm jet,49},
\end{split}
\end{equation}
where 
$D_{\rm L}=6.4\times10^{3}~{\rm Mpc}$ is the luminosity distance of PKS 0502+049 \citep{Drinkwater1997},
$f_{\rm d}=\ln(E_{\rm p,max}/E_{\rm p,min})\simeq 17$ with $E_{\rm p,max}\sim 3\times10^{16}{\rm ~eV}$ and $E_{\rm p,min}\sim 9\times10^{8}{\rm ~eV}$ as the maximum and minimum energies of the protons. 

We can evaluate the expected neutrino energy flux per flavor after considering neutrino oscillations, as 
 \begin{equation}
 \begin{split}
E_{\nu_\mu}^2\frac{dN_{\nu_\mu}}{dE_{\nu_\mu}} &\approx\frac{1}{6}f_{\rm pp}f_{\rm dis}f_{\rm c}P_{\rm jet}/(4\pi D_{L}^2f_{\rm d})\\
 &\simeq 1.3\times10^{-8}{\rm ~GeV ~cm^{-2} ~s^{-1}}~f_{\rm dis}P_{\rm jet,49}.
 \end{split}
 \end{equation}
We note that $\gamma$-rays with energy larger than a few ten GeV will be absorbed by $\gamma\gamma$ interactions with UV-optical photons produced in the clouds \citep{Poutanen2010, Stern2014nH}, while high-energy neutrinos can escape the source without attenuation.

\begin{figure}
\centering
\includegraphics[width=0.5\textwidth]{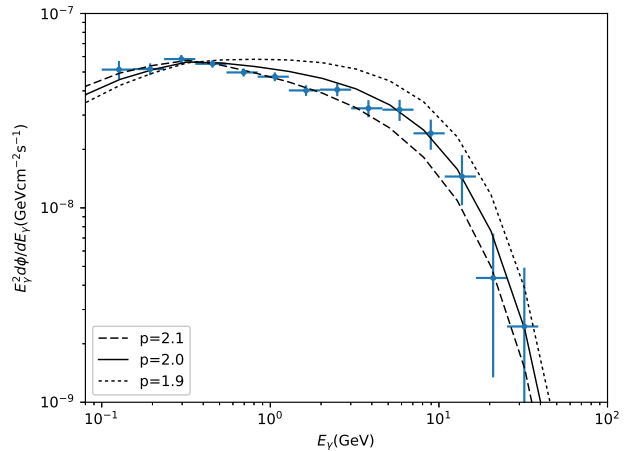}
\caption{$\gamma$-ray spectra of PKS 0502+049 during its active phase. The blue dots are the observed data from {\it Fermi}-LAT and the black lines are the prediction from the cloud-jet interaction model with different proton spectral indices as indicated in the legend.
}
\label{fig:gamma_spec}
\end{figure}

According to Eq. (\ref{F_gamma}), providing the observed $\gamma$-ray flux as in Figure \ref{fig:gamma_spec} via hadronuclear interactions
requires a high dissipation efficiency and a high jet power, i.e., $f_{\rm dis}P_{\rm jet,49}\sim 2$.

By assuming that both the observed GeV $\gamma$-rays and high-energy neutrinos originate from the same hadronuclear interactions, the measured flux of the former constrains the detection rates of the neutrinos.
Uncertainty in the spectral index of the accelerated protons leads to uncertainty in the neutrino flux. 

In Fig. \ref{fig:gamma_spec}, we plot three $\gamma$-ray spectra for different spectral indices $p$ to compare with observations.
We set the $\gamma$-ray flux at $0.5$ GeV to be the same value, by adjusting the jet power $P_{\rm jet}$.
For the remaining parameters, we fix $f_{\rm dis}= 1$, the observed cutoff energy of the $\gamma$-ray spectrum$E_{\gamma,\rm cut}=20~{\rm GeV}$, and the maximum energy of the accelerated protons $E_{\rm p,max}={\rm 30 ~{\rm PeV}}$.

\begin{figure}
\centering
\includegraphics[width=0.5\textwidth]{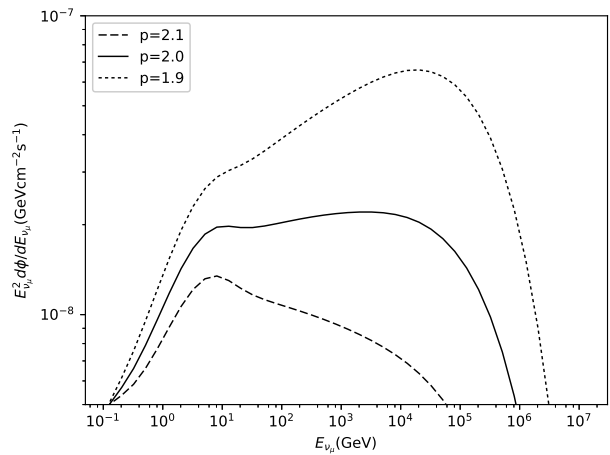}
\caption{Predicted muon neutrino spectra from the cloud-jet interaction model with the same parameters as in Figure \ref{fig:gamma_spec}.
}
\label{fig:numu_spec}
\end{figure}

Fig. \ref{fig:numu_spec} plots the corresponding muon neutrino spectra,  which can extend to energies $0.1E_{\rm p,max}/(1+z) \sim$ 150 TeV. 
Adopting the effective area for IceCube appropriate for the zenith angle of the object \citep{IceCube2018neutrinoflare}, we integrate the neutrino flux over the active duration of the blazar, and derive the detection rates of muon neutrinos by IceCube from PKS 0502+049.
The neutrino detection rates, the spectral index of accelerated protons and the required jet power for each model are listed in Table \ref{table}.

Table \ref{table} shows that with the constraint from the observed GeV $\gamma$-rays, our model can provide $4.5$ neutrinos for a flat spectrum of the accelerated protons and a jet power of $P_{\rm jet}=3.5\times10^{49}{\rm ~erg~s^{-1}}$.
More neutrinos can be produced for a harder proton spectrum and a higher jet power, for example, 15 neutrinos can be produced for $p=1.9$ and $P_{\rm jet}=7.2\times10^{49}{\rm ~erg~s^{-1}}$.

\begin{table}
\centering
\caption{\label{table} The predicted number of muon neutrinos detectable by IceCube from PKS 0502+049 during its active phase, for each proton spectral index corresponding to the three spectra in Figure \ref{fig:numu_spec}}
\begin{tabular}{ccc}
\hline
\hline
~~p~~&$P_{\rm jet}$&$~~N_{\nu_\mu}~~$\\
&$~~~(10^{49}{\rm ~erg\ s^{-1}})~~~$&\\
\hline
$1.9$& 7.2& 15 \\ 
$2.0$ &3.5 &4.5\\   
$2.1$ &2.2&1.4 \\
\hline
\end{tabular}
\end{table}

{\it Discussion:}
In this work, 
we assume that high-energy protons are confined in the clouds during the time that the clouds cross the jet.
Thus the efficiency of hadronuclear interactions can reach unity, since the $pp$ cooling timescale $t_{\rm pp}$ is much smaller than the crossing timescale of the cloud $t_{\rm cr}$.

The protons confined in the cloud interact with the dense gas in the cloud and emit $\gamma$-rays and high-energy neutrinos isotropically in the rest frame of the cloud.
The motion of the cloud along our line of sight is non-relativistic, implying no significant Doppler boosting of the emitted photons and neutrinos \citep{Barkov2012JetCloudinteraction}.
An expected spectral feature is the pion bump at $\sim 67/(1+z){\rm ~MeV}$ in the $\gamma$-ray and neutrino spectra that reflect the threshold energy of $pp$ collisions, of which future searches in $\gamma$-rays will help us to confirm or exclude a hadronuclear origin.

The mass of the central black hole for PKS 0502+049 is $7.5\times10^{8}M_\odot$ \citep{Oshlack2002BHmass},
with corresponding Eddington luminosity $L_{\rm Edd}=9.5\times10^{46}{\rm ~erg~s^{-1}}$.
To explain the gamma-ray flares and neutrino flare, the required accretion rate of PKS 0502+049 must exceed the Eddington luminosity by about two orders of magnitude during the flaring phase.
Though it might be rare, the super-Eddington case can be achieved for jets from tidal disruption of massive stars \citep{Wu2018super-eddington}.

The rate of observing such an association between neutrino flares and $\gamma$-ray flares in FSRQs depends mainly on the distance of the object, the frequency of clusters of clouds crossing the jet, the power of the central engine during the flares, the dissipation efficiency, the efficiency that the accelerated protons reach the cloud, and the hadronuclear interaction efficiency, most of which have large uncertainties.
Their effect on the observables are embodied together in the rate of such flares and the $\gamma$-ray flux of the flares, if the $\gamma$-rays arise from pion decay in hadronuclear interactions.

The high-energy $\gamma$-ray photons are likely attenuated by $\gamma\gamma$ pair production with UV-optical photons from the clouds \citep{Poutanen2010, Stern2014nH}, as well as cosmic optical and infrared background photons during their intergalactic propagation \citep{YInoue2013_EBL}, which is consistent with the fact that most FSRQs have soft $\gamma$-ray spectra \citep{Liao2018}.
Therefore, the intrinsic $\gamma$-ray photon fluence before attenuation is expected to be higher than the observed fluence.
The value of the intrinsic fluence depends on the spectral index of the accelerated protons, as well as details of the UV-optical spectra of the cloud and the extragalactic background light. 
FSRQ flares with a high intrinsic $\gamma$-ray photon fluence are promising candidates for detecting correlated neutrino flares.
Future detailed studies of the long-term $\gamma$-ray variability of FSRQs will help us to understand better the probability of detecting associated $\gamma$-ray and neutrino flares.

Note that the duration of the $\gamma$-ray flares also affects the detectability, since a longer duration will imply more background events for neutrinos.
Another important aspect is the sky location of the targets with respect to IceCube, since it is most sensitive to high-energy neutrinos at zenith angles slighly below the horizon at the Sourth Pole
\citep{IceCube2018neutrinoflare}.

Both time-integrated \citep{Braun2008pointsourceneutrino} and time-dependent \citep{Braun2010Time-dependent} searches in IceCube historical as discussed in \citep{IceCube2015timedependent,IceCube2017timeintegrated,IceCube2018neutrinoflare} for further flares with high intrinsic $\gamma$-ray fluence in FSRQs including PKS 0502+049 are highly motivated.
With more detections of the association between high-energy neutrinos and the active state of FSRQs in the future, the uncertainties of
the cloud-jet model can be further constrained, e.g. cloud density, jet power, dissipation efficiency, and confinement of protons in the cloud.

\acknowledgments
We thank Shigehiro Nagataki, Hirotaka Ito and Ruoyu Liu for useful discussions. H.N.H. is supported by National Natural Science of China under grant 11303098, and the Special Postdoctoral Researchers (SPDR) Program in RIKEN. YI is supported by RIKEN iTHEMS Program, JSPS KAKENHI Grant Number JP16K13813, and  Leading Initiative for Excellent Young Researchers, MEXT, Japan. SI thanks support from JSPS KAKENHI Grant Number JP17K05460.

\end{document}